\documentclass[10pt,a4paper,twocolumn]{article}

\usepackage{amssymb}
\usepackage{amsmath}
\usepackage{amsfonts}
\usepackage{graphicx}
\usepackage{epstopdf}
\usepackage{cite}
\usepackage[title]{appendix}
\usepackage[parfill]{parskip}
\usepackage{setspace}
\usepackage[font=singlespacing]{caption}
\usepackage{xcolor}
\usepackage{soul}

\usepackage{authblk}
\usepackage{sectsty}
\sectionfont{\bfseries\Large\raggedright}

\def\lsim {\ifmmode {\buildrel<\over\sim}}

\def\sss{\scriptscriptstyle\rm}

\def\1var{(\bx_1...\bx\N)}

\def\br{{\bf r}}

\def\b1{{\bf 1}}
\def\bx{{x}}

\def\N{_{\sss N}}

\def\pr{^{\prime}}

\def\ua{_{\alpha}}
\def\vp{v_{R}}
\def\uat{_{atom}}
\def\ume{_{metal}}

\usepackage{epstopdf}
\usepackage[caption=false]{subfig}

\usepackage[numbers,sort&compress,merge]{natbib}
\bibpunct[, ]{[}{]}{,}{n}{,}{,}

\begin{document}
\date{}

\title{Chemical Potential of Integer Electron Systems \\ (submitted to Molecular Physics)}

\author[1]{K. Niffenegger}
\author[2]{Y. Oueis}
\author[2]{J. Nafziger}
\author[1,2]{Adam Wasserman}

\affil[1]{Department of Physics and Astronomy, Purdue University, 525 Northwestern Avenue, West Lafayette, Indiana 47907, USA}
\affil[2]{Department of Chemistry, Purdue University,  560 Oval Drive, West Lafayette, Indiana 47907, USA}

\maketitle

\begin{abstract}
A truly isolated atom always has an integer number of electrons. If placed in contact with a far-away metallic reservoir, a {\em range} of metallic chemical potentials $\mu$ will lead to an identical number of electrons, $N$, on the atom.  We formulate a density embedding method in which the range of $\mu$ leading to integer $N$ decreases due to finite-distance interactions between the metal and the atom. The typical $N(\mu)$ staircase function is smoothed out due to these finite-distance interactions, resembling finite-temperature effects. Fractional occupations on the atom occur only for sharply-defined $\mu$'s. We illustrate the new method with the simplest model system designed to mimic an atom near a metal surface. Because calculating fractional charges is important in various fields, from electrolysis to catalysis, solar cells and organic electronics, we anticipate several potential uses of the proposed approach.
\end{abstract}

\section{Introduction}  %

\begin{figure}
\centering
\includegraphics[width=3in]{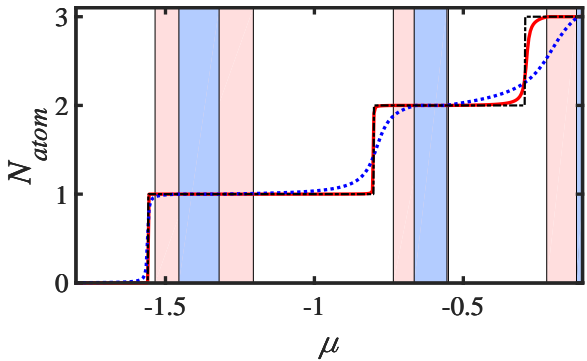}
\caption{The atomic fragment occupation number $N\uat$ as a function of the system chemical potential, $\mu$, for $R =3$ (dotted blue line), $R = 5$ (solid red line), and $R > 10$ (dot-dash black line). The step-like behavior that occurs at large separations smooths out as we bring the fragments closer together. The light blue and salmon-shaded regions highlight $\mu$ values for which $N\uat$ is exactly integer at $R =3$ and $R=5$ respectively.}
\label{fig:muvNRall}
\end{figure}

When the Hohenberg-Kohn theorem \cite{HK} was extended to fractional electron numbers by Perdew, Parr, Levy, and Balduz (PPLB, \cite{PPLB,Perdew85,PL84}), a result of far-reaching consequences was found: The ground-state energy of an $N$-electron system, $E(N)$, is a piecewise-continuous linear function of $N$:
\begin{equation}
E(N) = (1 - \omega)E(m) + \omega E(m + 1)~~,
\label{e:PPLB}
\end{equation}
where $E(m)$ and $E(m + 1)$ are the ground-state energies of the (integer) $m$ and $(m+1)$-electron systems, and $0\leq \omega \leq 1$.  At strictly zero temperature, an atom or molecule that is in equilibrium with a \textit{far-away} metal reservoir, will be neutral in the ground state for any chemical potential $\mu$ in the range 
\begin{equation}
\label{e:PPLBmu}
-I < \mu < -A~~,
\end{equation}
where $I$ is the ionization potential and $A$ the (positive) electron affinity of the neutral atom. 
For chemical potentials lower than $-I$, the atom transfers one electron to the reservoir. For chemical potentials higher than $-A$, the atom receives one electron from the reservoir.  The number of electrons in the atom is thus a staircase function of the chemical potential (black dot-dash line in Fig. \ref{fig:muvNRall}), which is clearly only sharply defined for non-integer numbers. The range of $\mu$ that is consistent with the integer $m$ is the fundamental energy gap of the atom, $E_g = I-A$, which is thus given by the total discontinuity in the derivative of $E(N)$ with respect to $N$ at $N=m$. All properties of the system involving derivatives of the energy with respect to $N$ are similarly undefined at the integers at zero temperature.

A smoothening of the discontinuities at integer numbers of electrons and a range of $\mu$ that is narrower than $I - A$ can be found by applying techniques of the grand-canonical ensemble at finite temperature \cite{GH68}. The main result of our work is that sharper values of $\mu$ can be found even at zero temperature by considering \textit{finite distances} from the metal reservoir. To show this, an unambiguous definition is needed for the charge of the atom when it is located at an interacting distance from the metal. We provide such definition by requiring that the chemical potential of the two fragments (metal and atom) be equal while satisfying the standard constraint of density-embedding methods, i.e. that the sum of the two fragment densities be equal to the total electronic density. With this definition of fragments, the regions of strictly integer numbers of electrons on the atom are narrower than $I-A$ when the atom is at an interacting distance from the reservoir (red and blue lines in Fig. \ref{fig:muvNRall}). Outside of the shaded regions in Fig. \ref{fig:muvNRall}, the atom acquires a fractional number of electrons. At large separations between the atom and the metal, our model recovers the PPLB results. 
At shorter separations, the regions of integer occupations shrink but do \textit{not} collapse to a single point. Due to the finite-distance interactions, the effective values of $I$ and $A$ are different from those of the isolated atom. As a result, the narrowing of the integer windows is not symmetric with respect to ${(I+A)}/{2}$ and is markedly different near different integer occupations.
 
The method is described in Sec. \ref{sec:CPCPDFT} and illustrated through explicit numerical computation in Sec. \ref{sec:illustr}. We end with a brief summary and outlook in Sec. \ref{sec:outlook}.

\section{Chemical-potential constrained Partition-DFT}
\label{sec:CPCPDFT}

Consider a system of electrons in an external potential $v(\br)$ that can be written as:
\begin{equation}
v(\br)=v_{atom}(\br)+v_{metal}(\br)~~,
\label{e:v_tot}
\end{equation}
where $v_{metal}(\br)$ describes a background periodic or semi-periodic metallic potential supporting a continuum of electronic levels occupied up to a Fermi energy, $\epsilon_F$, and  $v_{atom}(\br)$ is a localized potential such as the Coulomb or screened-Coulomb potential of an atom. The partition of Eq.(\ref{e:v_tot}) is useful when one wants to describe an atomic defect in a solid or an atom adsorbed on a metal surface. 

The task of finding the number of electrons on the atom, $N_{atom}$, is non-trivial unless $v_{atom}(\br)$ is far from all regions where $v_{metal}(\br)$ is non-zero, in which case one recovers the black staircase function of Fig.1 with $\mu=\epsilon_F$. 
The total density $n(\br)$ for the combined system of atom and metal can be partitioned as $n_{atom}(\br)+n_{metal}(\br)$ in many different ways. Partition Density Functional Theory (P-DFT, \cite{CW07,EBCW10,NW14}) provides an elegant, unambiguous method for performing such a partition when the number of electrons is finite and the external potential for each fragment vanishes in all directions as $|\br|\to\infty$. Fragments in P-DFT are isolated from each other and are in contact with a far-away electronic reservoir through which they can exchange electrons. The interaction energy between the fragments is recovered by means of a unique global embedding potential, referred to here as the reactivity potential, $v_R(\br)$. The prescription to determine $N_{atom}$ becomes simple: Minimize the sum of the fragment energies (i.e. atom and metal) subject to the constraint that the fragment densities sum to to the correct total density, and then calculate the number of electrons in the atom as $N_{atom}=\int n_{atom}(\br) d\br$. This number is in general not an integer because each P-DFT fragment energy is given by the ensemble expression of Eq.(\ref{e:PPLB}), where the non-integer $\omega$ is one of the parameters to be optimized during the energy minimization.

In the case of the potential of Eq.(\ref{e:v_tot}), however, $v_{metal}(\br)$ does not vanish as $|\br|\to\infty$ in all directions, and one of the fragment energies is infinite.  The approach of P-DFT is thus not directly applicable. 

In lieu of an energy minimization, we propose here to impose a chemical-potential equalization constraint, shown to be equivalent to energy-minimization for the case of finite systems \cite{CW06}. The prescription is just as simple: Find the fragment densities that equalize the chemical potentials of the fragments and the chemical potential of the combined system:
\begin{equation}
\label{eq:mucond}
\mu_{atom}=\mu_{metal}=\mu~~,
\end{equation}
while adding to the correct total density. The resulting density of the atom is an ensemble ground-state density of $v_{atom}$ that is modified by the addition of $v_R(\br)$, which is identical for both atomic and metallic fragments.

When $N_{atom}$ is an integer, $\mu\uat$ is defined only within a range, so Eq. \ref{eq:mucond} is applicable only for \textit{non}-integer values of $N_{atom}$. For integer occupations, the condition of Eq. \ref{eq:mucond} is modified taking into account Eq. \ref{e:PPLBmu}:
\begin{equation}
-I_{atom} < \mu_{metal}=\mu <  -A_{atom}~~,
\label{eq:muRange}
\end{equation}
where $I_{atom}$ and $A_{atom}$ are computed {\em in the presence of} $\vp(\br)$. We consider our method converged if either $N_{atom}$ is non-integer and condition \ref{eq:mucond} is satisfied \textit{or} if $N_{atom}$ is integer and condition \ref{eq:muRange} is satisfied. In the following section, we successfully apply this method to a model system that mimics an atom-metal interface in 1-D; however, the rigorous derivation of the conditions for the existence of a unique reactivity potential for systems with semi-infinite fragments is still not established.
	
\section{Simple Illustration}  %
\label{sec:illustr}

We choose the simplest non-trivial system that exhibits the features we need: One semi-infinite fragment (the `metal') and one finite fragment with a small number of bound states (the `atom'). The total number of electrons is infinite, but the electrons are non-interacting and restricted to move in only one dimension. 

\begin{figure}
\centering
	\includegraphics[width=3in]{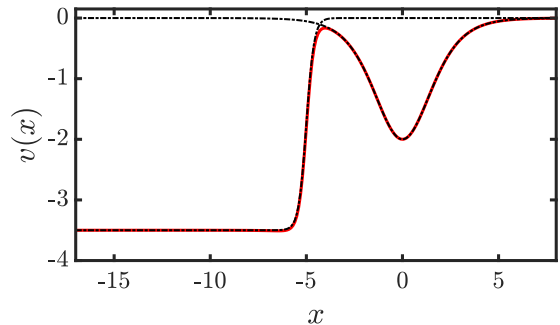}
	\caption{The potentials $v_{metal}$ and $v_{atom}$ (dashed black lines) along with the total external potential $v$ (solid red line) for the parameters $R = 5$, $\gamma = 0.5$, $Z = 2$, and $V_{0} = 3.5$.}
	\label{fig:Vext}
\end{figure}
%

\subsection{Model System}  

The metal is represented by a potential that goes to a negative constant $-V_0$ as $x\to -\infty$:
\begin{equation}
v_{metal}(x) =\frac{-V_0}{1+e^{s(x-R)}}~~,
\label{v-metal}
\end{equation}
and is populated with non-interacting `spinless electrons' up to the Fermi level $\epsilon_F$, with $-V_0<\epsilon_F<0$. In Eq. \ref{v-metal}, $R$ is the separation between the metal surface and the center of the atomic potential, and $s$ is a parameter that determines the steepness of the step. The form of the potential allows it to be smooth enough to be used with finite-difference methods on a spatial grid while preserving a steep step-like feature. The atom is represented by a finite potential with a finite number of bound states:
\begin{equation}
v_{atom}(x) = -Z\mathrm{cosh}^{-2}(\gamma x)~~,
\end{equation}
where $Z$ and $\gamma$ are parameters that control the depth and width of the well. We use $V_{0}=3.5$, $Z=2$, and $\gamma=0.5$ throughout the paper. The total external potential is then just the sum of $v_{metal}$ and $v_{atom}$ according to Eq.(\ref{e:v_tot}), as shown in Fig. \ref{fig:Vext}.
\begin{figure}
\centering
\includegraphics[width=3in]{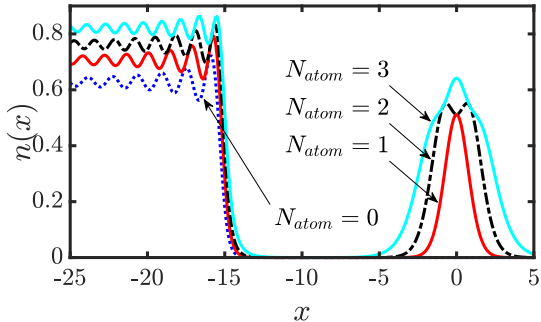}
\caption{Total system densities, $n(x)$, for four choices of $\mu$ which give $N\uat = 0, 1, 2,$ and $3$ using a separation $R = 15$. }
\label{fig:IntNRefDens}
\end{figure} 

The full system of metal plus atom produces a continuum of states. The chemical potential of this system is its Fermi energy. The reactivity potential $\vp(x)$ ensures fragment densities sum to the total density of the system. The densities of the total system and of the metal fragment are calculated as the integral \cite{BOOK,Near1D}:
\begin{equation}
n(x) = \frac{1}{2\pi i} \int_{C_{\mu}} G_E (x,x) dE~~. 
\label{GDensCalc}
\end{equation}
Here, $G_E(x,x^\prime)$ is the corresponding Green's function that can be found numerically exactly. The integral is evaluated over $C_{\mu}$, a contour in the complex energy plane containing all possible occupied states. The total system density $n(x)$ for a large separation $R = 15$ is shown in Fig. \ref{fig:IntNRefDens}. We can see that as the chemical potential of the system increases through the energy levels of the isolated atomic potential, the density near the atom increases in large jumps every time the chemical potential reaches a bound state. The atomic densities have the ensemble form \cite{PPLB}:
\begin{equation}
\label{nEns}
n\uat(x)=\omega n_{p+1}(x)+(1-\omega)n_{p}(x)~~,
\end{equation}
where $p$ is the lower bounding integer of $N\uat$, $0\leq \omega <1$, and $N\uat=p+\omega$. Calculations of the atomic densities at integer occupations are trivial.

The eigenvalues of $v_{atom}(x)$, when isolated, are known for any chosen $Z$ and $\gamma$ \cite{LLQM}:
\begin{equation}
\label{LLEig}
\varepsilon_{i}^{(0)} = -\frac{\gamma^{2}}{8}\Big\{-[1+2(i-1)]+\sqrt{1+(8Z\gamma^{2})}\Big\}^{2}~~,
\end{equation}
where $i$ runs from 0 to the maximum number of bound states. The superscript `$(0)$' indicates that the atom does not interact with the metal. These eigenvalues broaden into resonances when the atom couples to the metal and $|\epsilon_F|>|{\varepsilon_i}|$.

\subsection{Search for chemical-potential Equalization}    %

To obtain a single point on the $N_{atom}$ versus $\mu$ plot in Fig. \ref{fig:muvNRall}, we perform a numerical algorithm for a set value of $\mu$. This algorithm consists of the `inner' inversion that computes the reactivity potential at the current guess of $N_{atom}$ and the `outer' loop that updates $N_{atom}$ until one of the chemical-potential equalization conditions, Eq. \ref{eq:mucond} or Eq. \ref{eq:muRange}, is satisfied. Our inversion method requires the precomputed total density $n(x)$ for each $\mu$. We set $\mu_{metal}$ equal to $\mu$ and do not vary it throughout the inversion procedure.
\begin{figure}
\centering
\includegraphics[width=3in]{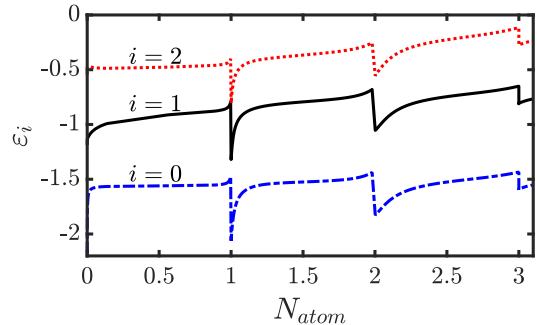}
\caption{The atomic fragment eigenvalues $\varepsilon_{i}$ as the fragment occupation number $N\uat$ passes through the integer occupation of one for $R = 3$. }
\label{fig:Evals_2cosh5_R3}
\end{figure} 

We choose $v_{R,Guess}^{(0)}(x)=0$ as our initial guess for $\vp(x)$. To calculate the initial guess for $N_{atom}$, we start by calculating the isolated atomic density, $n_{atom}^{(0)}(x)$. We separate $n_{atom}^{(0)}(x)$ into contributions from the density of the highest occupied atomic orbital (HOMO) and the density due to the core electrons, $n_{atom}^{(0)}(x) = n_{core}^{(0)}(x)+n_{\rm HOMO}^{(0)}(x)$. The number of states included in the core region, $N_{core}$, is equal to the number of eigenvalues of the isolated atom which are below $\mu$. The initial guess for the atomic occupation is $N_{atom}^{(0)}=N_{core}+\omega^{(0)}$, where $\omega^{(0)}$ is found as:
\begin{equation}
\omega^{(0)} =\int dx \Big[n(x)-n_{metal}^{(0)}(x)-n_{core}^{(0)}(x)\Big]~~,
\end{equation}
and $n_{metal}^{(0)}(x)$ is the density of the isolated metal fragment.

At each iteration $i \geq 0$ of the  `outer' loop, we use the current values $N\uat^{(i)}$ and $v_{R,Guess}^{(i)}(x)$ to compute the $\vp^{(i)}(x)$ that minimizes the difference between $n(x)$ and $n_{f}^{(i)}(x)=n_{metal}^{(i)}(x)+n_{atom}^{(i)}(x)$ to numerical precision (i.e. $v_{R,Guess}^{(i)}(x)$ is used as an initial guess to find $\vp^{(i)}(x)$ at fixed $N\uat^{(i)}$ and $\mu_{metal}$). The resulting fragment densities are used to calculate the fragment responses $\chi\ua^{(i)}(x,x\pr)$ that will be used to update $v_{R,Guess}^{(i)}(x)$:
\begin{equation}
\label{Resp}
\chi\ua^{(i)} (x,x\pr) = \frac{dn\ua^{(i)}(x)}{dv_{eff,\alpha}^{(i)}(x\pr) }~~,
\end{equation} 
where $v_{eff,\alpha}^{(i)}(x) = v\ua(x) + \vp^{(i)}(x)$ and $\alpha$ is either `metal' or `atom.'

If $N\uat^{(i)}$ is not an integer, then $\mu\uat^{(i)}$ equals the HOMO energy $\varepsilon_{\rm HOMO}^{(i)}$ in the presence of $\vp^{(i)}(x)$, and we use Eq. \ref{eq:mucond} to check if the algorithm has converged. On the other hand, if $N\uat^{(i)}$ is an integer, we use the convergence criteria of Eq. \ref{eq:muRange}, with $-I_{atom}=\varepsilon_{\rm HOMO}^{(i)}$ and $-A_{atom}=\varepsilon_{\rm LUMO}^{(i)}$, where $\varepsilon_{\rm LUMO}^{(i)}$  is the energy of lowest unoccupied atomic orbital in the presence of $\vp^{(i)}(x)$.

If neither of the conditions is met, we continue by calculating $N\uat^{(i+1)}$ as:
\begin{equation}
\label{e:Nupdate}
N\uat^{(i+1)}=N\uat^{(i)} + \frac{\mu_{metal}^{(i)} - \mu\uat^{(i)}}{{d\mu\uat^{(i)}}/{dN\uat^{(i)}}}~~,
\end{equation}
where:
\begin{equation}
\label{dMudN}
\frac{d\mu\uat^{(i)}}{dN\uat^{(i)}} = \int dx n_{\rm HOMO}^{(i)}(x) \frac{dv_{R}^{(i)}(x)}{dN\uat^{(i)}}~~.
\end{equation}
In Eq. \ref{dMudN}, the derivative on the right hand side is given by:
\begin{equation}
\label{dVpdN}
\begin{split}
\frac{dv_{R}^{(i)}(x)}{dN\uat^{(i)}}&=\int dx\pr n_{\rm HOMO}^{(i)}(x)\Big[\Big(\chi_{metal}^{(i)}(x,x\pr)\Big)^{-1} +\\
&\Big(\chi\uat^{(i)}(x,x\pr)\Big)^{-1}\Big]~~.
\end{split}
\end{equation}
Eq. \ref{dVpdN} is also used to update the guess $v_{R,Guess}^{(i+1)}(x)=v_{R}^{(i)}(x) +dv_{R}^{(i)}(x)$.

\subsection{Energies, densities, and reactivity potentials}  %

The origin of the discontinuities of the chemical potential can be understood in terms of the atomic orbitals $\varepsilon_i$ (in the presence of $\vp(x)$). Near integer occupations, the energy of the HOMO shifts up from the left and the energy of the LUMO shifts down from the right, as we see in Fig. \ref{fig:Evals_2cosh5_R3}. Even for separations as small as $R = 3$, levels do not equalize.

\begin{figure}
\centering
\includegraphics[width=3in]{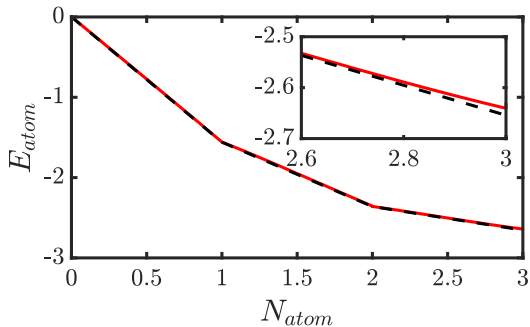}
\caption{The atomic fragment energy $E\uat$ as a function of the fragment occupation number $N\uat$ for $R = 3$ (solid red line) and $R = \infty$ (dashed black line). }
\label{fig:EvNall}
\end{figure}

The effect of the finite-distance interactions on the energy of the atom can be seen in Fig. \ref{fig:EvNall}. The energy of the atom is defined as the sum of occupied orbitals \textit{minus} the energy contribution from the reactivity potential:
\begin{equation}
\label{eq:Eat}
\begin{split}
E\uat &\equiv\sum_{i=1}^{p\uat}\varepsilon\uat^{i} + w\varepsilon\uat^{\rm LUMO} \\
&-\int dx \vp(x) n\uat(x)~~. 
\end{split}
\end{equation}
In Fig. \ref{fig:EvNall}, the dashed line shows the energy at large separation, $R=15$. It consists of \textit{straight} line segments \cite{PPLB,Perdew85,PL84}. At short distances (e.g. $R = 3$, solid red in Fig. \ref{fig:EvNall}) the line segments have a slight curvature. As shown in the inset plot of Fig. \ref{fig:EvNall}, the curvature is more noticeable for $N\uat$ in the range of $2$ to $3$, where $E\uat$ values are more evenly spaced. This curvature is the consequence of the inter-fragment interactions, but it does not smoothen the cusps at integer occupations.

\begin{figure}
\centering
  \begin{tabular}{@{}c@{}}
  \includegraphics[width=3in]{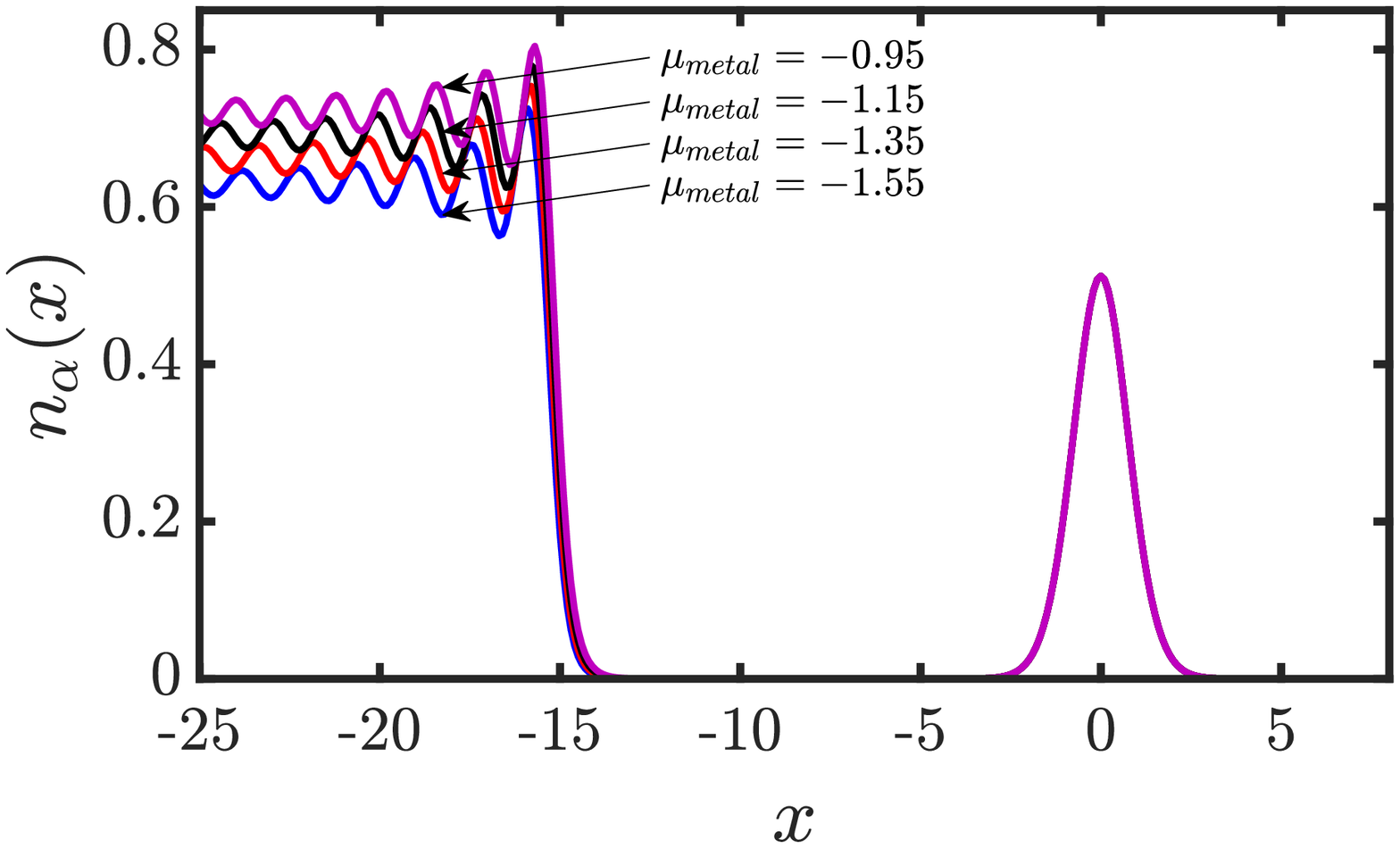} \\
  \includegraphics[width=3in]{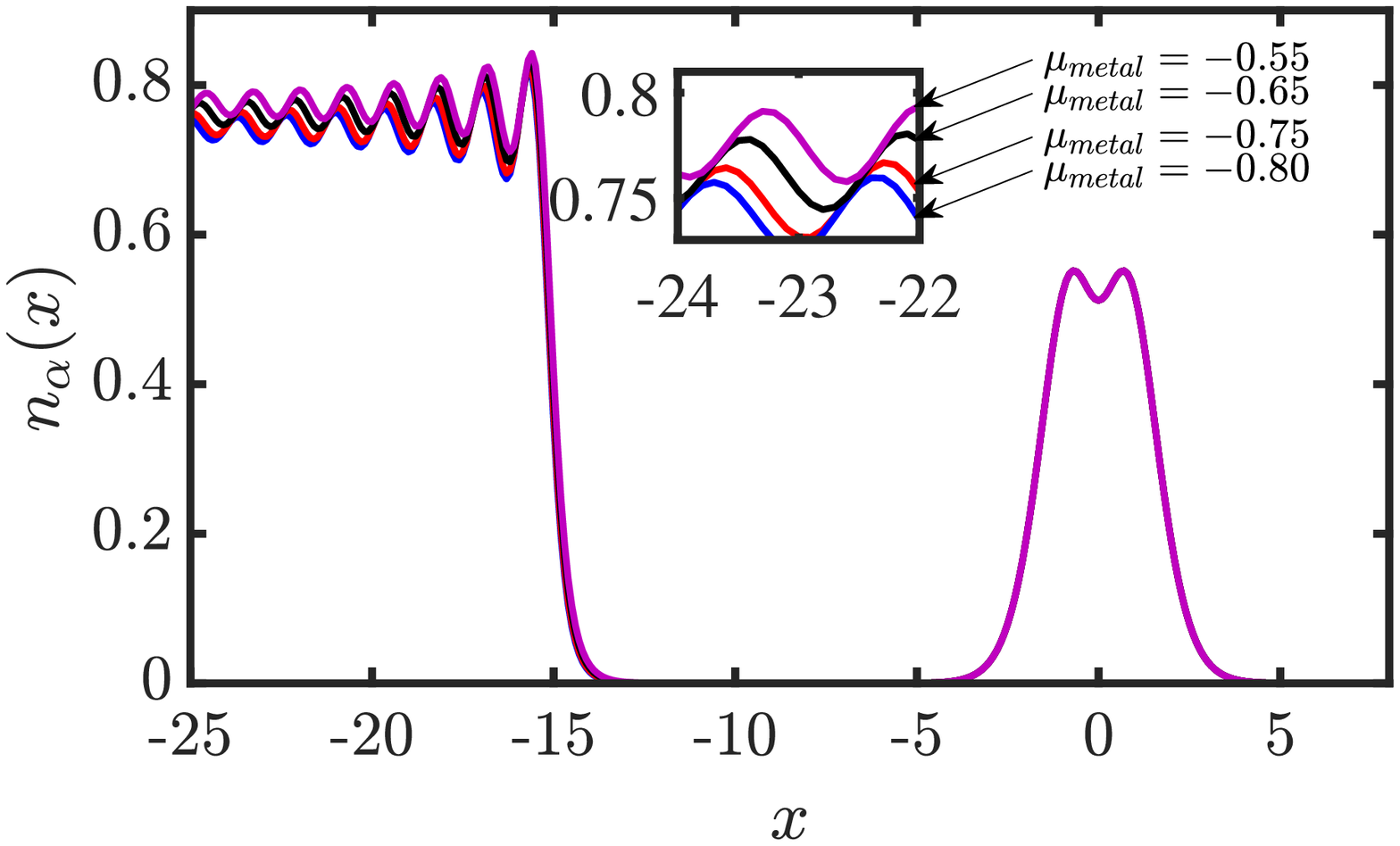} \\
  \includegraphics[width=3in]{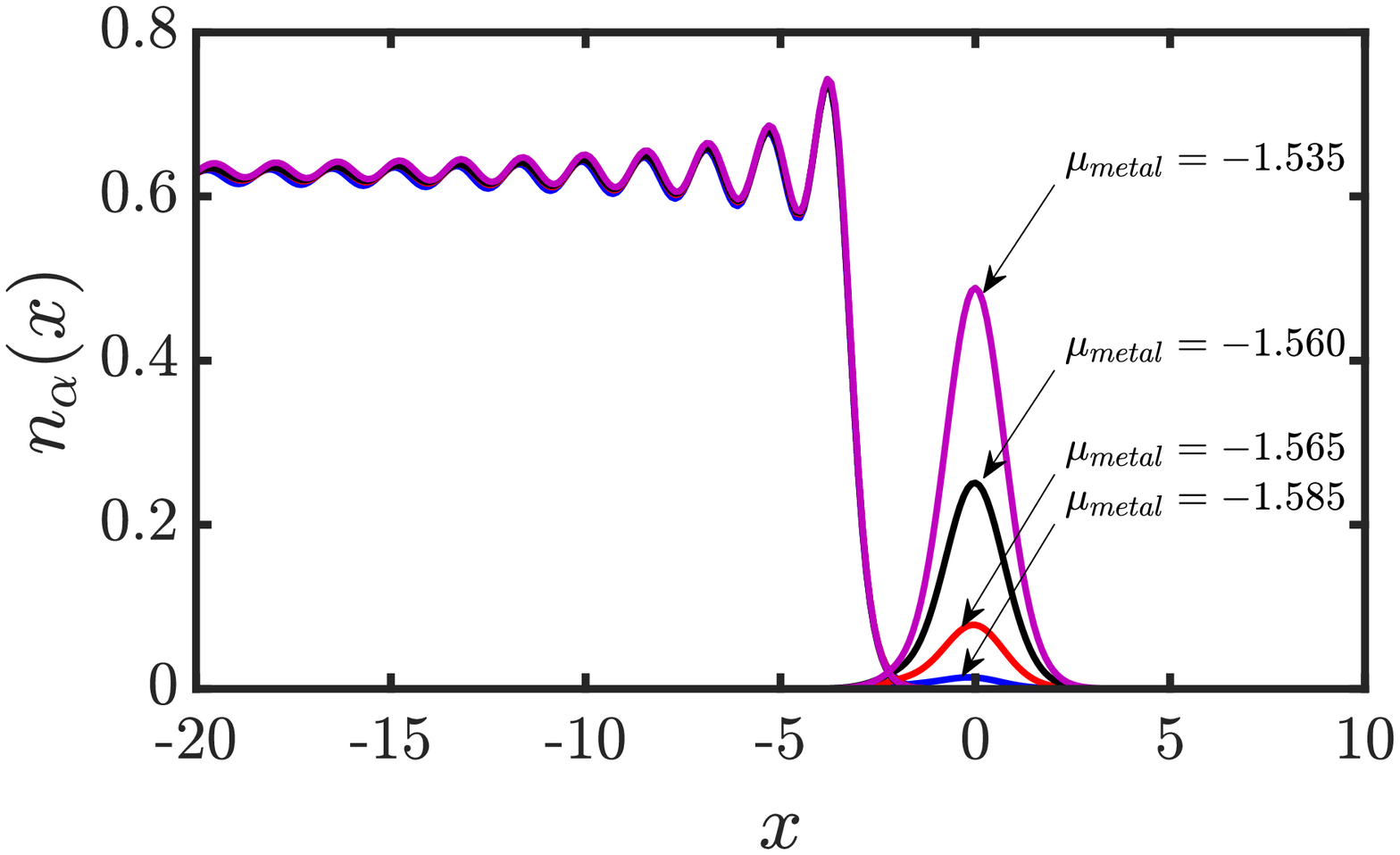}
  \end{tabular}
  \caption{Fragment densities $n\ua(x)$ at various values of $R$ and $\mu\ume$. Top: $R=15$ and values of $\mu\ume$ corresponding to $N\uat = 1$: $-1.55$ (blue), $-1.35$ (red), $-1.15$ (black), and $-0.95$ (violet). Middle: $R=15$ and values of $\mu\ume$ corresponding to $N\uat = 2$: $-0.8$ (blue), $-0.75$ (red), $-0.65$ (black), and $-0.55$ (violet). Bottom: $R = 3$ and values of $\mu\ume$ producing values of $N\uat$ between $0$ and $1$: $-1.585$ (blue), $-1.565$ (red), $-1.56$ (black), and $-1.535$ (violet).}
\label{fig:FragDens}
\end{figure}

The atomic fragment density at large values of $R$ jumps abruptly when going through integer occupations, as can been seen in the top ($N\uat=1$) and middle ($N\uat=2$) panels of Fig. \ref{fig:FragDens}. For each value of $N\uat$, increasing the Fermi energy of the system changes almost  exclusively $n\ume(x)$. As these changes occur, we observe an increase in the value of the metal density accompanied by a decrease in the period of density (Friedel) oscillations. The bottom panel of Fig. \ref{fig:FragDens} shows the representative behavior of fragment densities at small separations. Densities corresponding to non-integer values of $N\uat$ begin to appear. We note that, in this regime, the density of the metal fragment appears unchanged for different values of $\mu\ume$. The density response of the system to infinitesimal changes of $\mu$ is thus largely localized to either atom or metal fragments.

\begin{figure}
\centering
\includegraphics[width=3in]{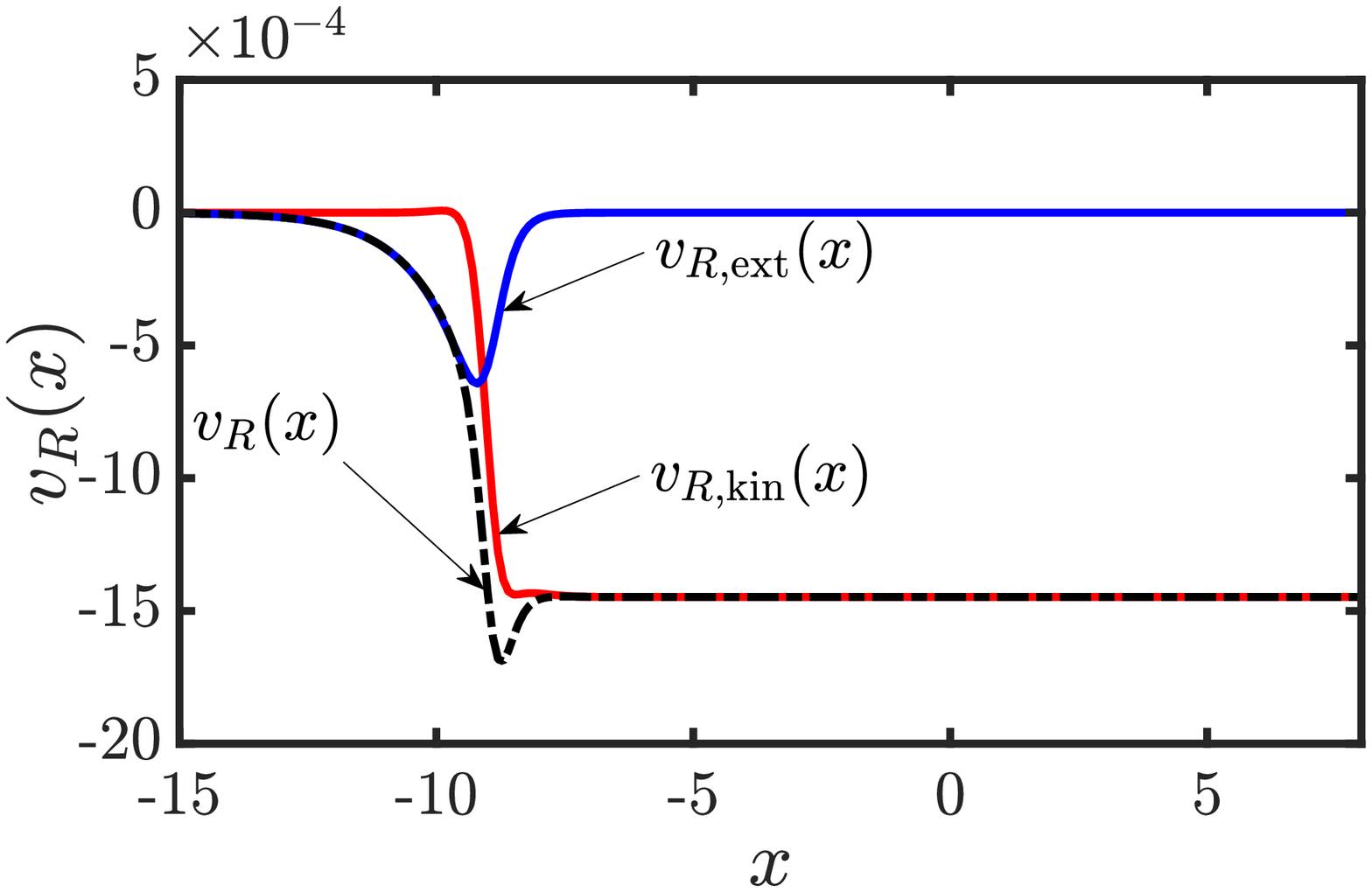}
\caption{The reactivity potential (black) along its components $v_{R,\mathrm{ext}}(x)$ (blue) and $v_{R,\mathrm{kin}}(x)$ (red) at $R = 15$ and $\mu\ume$ corresponding to $N\uat = 2$.}
\label{fig:vp}
\end{figure}

Finally, we point out a connection between the features of $\vp(x)$ and the known features of {\em exact} Kohn-Sham (KS) potentials, $v_{s}(x)$, at interfaces \cite{HK,Perdew85,KSStepsDD,KSPotentialFrac,ExactKSPotential,TMM09,GB96}. 
Since we work with non-interacting electrons, $\vp(x)$ has contributions only from the non-additive external potential ($v_{R,\mathrm{ext}}$) and from the non-additive kinetic potential ($v_{R,\mathrm{kin}}$):
\begin{equation}
\vp(x) = v_{R,\mathrm{ext}}(x) + v_{R,\mathrm{kin}}(x),
\label{e:v_R}
\end{equation}
where $v_{R,\mathrm{ext}}(x)$ is given by:
\begin{equation}
\label{eq:vpext}
\frac{n\ume(x)}{n(x)}v\uat(x) +\frac{n\uat(x)}{n(x)}v\ume(x),
\end{equation}
and $v_{R,\mathrm{kin}}(x)=v_R(x) - v_{R,\mathrm{ext}}(x)$.
%
%
In Fig. \ref{fig:vp}, we plot $\vp(x)$ and its components at large inter-fragment separation when $N\uat=2$. We observe that $v_{R,\mathrm{ext}}(x)$ has a well in the low density region. In contrast, $v_{R,\mathrm{kin}}(x)$ has a step-like feature analogous to the feature known to be present in $v_{s}(x)$ when two inequivalent fragments are separated (note from Eqs. \ref{e:v_R}-\ref{eq:vpext} that $v_{R,\mathrm{kin}}(x)=v_s[n\ume](x){n\ume(x)}/{n(x)}+v_s[n\uat](x){n\uat(x)}/{n(x)}-v_s[n](x)$). The magnitude of the feature in $v_{R,\mathrm{kin}}(x)$ highlights the importance of non-additive non-interacting kinetic energy functional approximations for practical embedding calculations. \cite{NJW17,JNW18} 

\subsection {Finite Distance as Finite Temperature}  %

\begin{figure}[p]
\centering
  \begin{tabular}{cc}
  \includegraphics[width=3in]{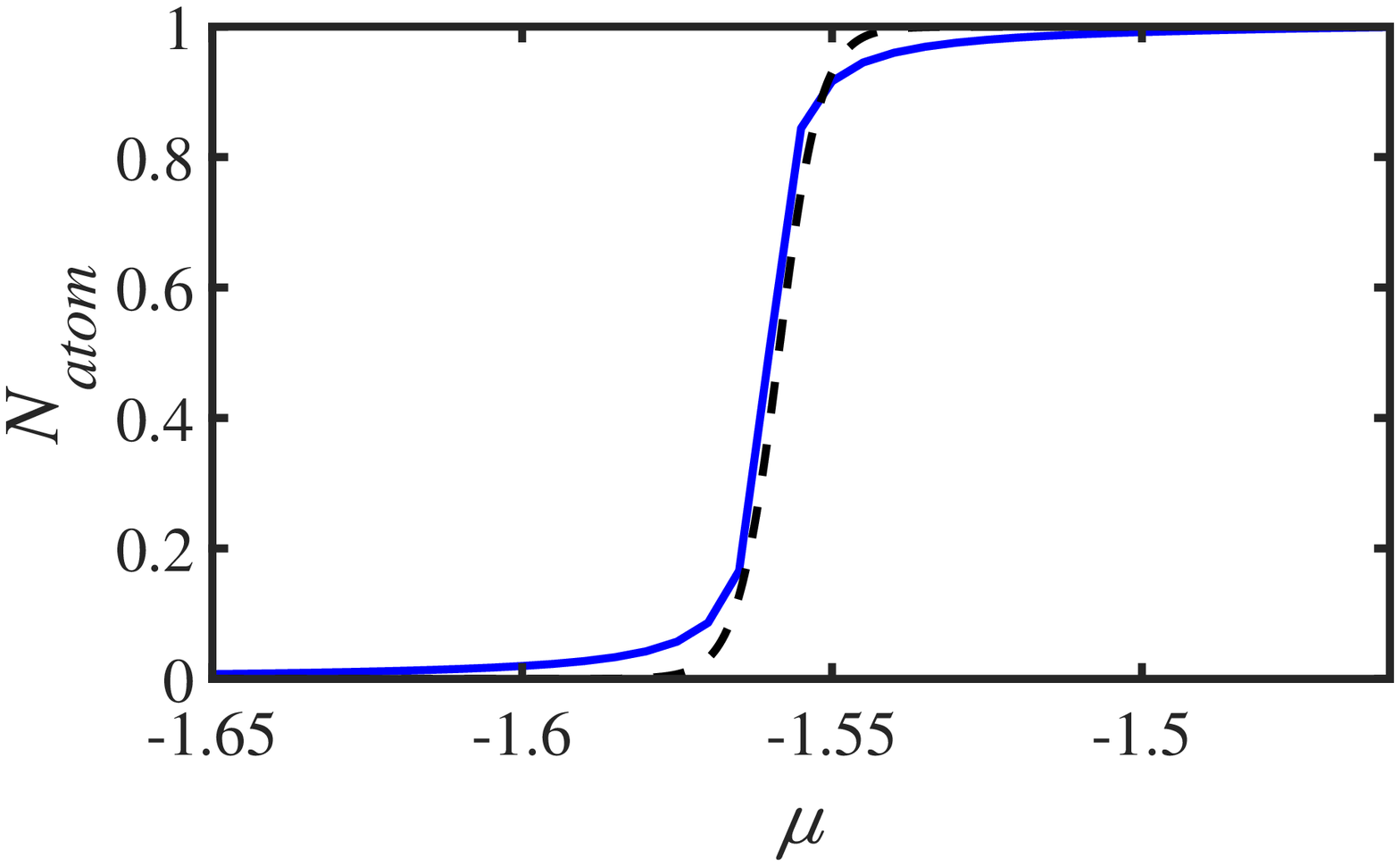} \\
  \includegraphics[width=3in]{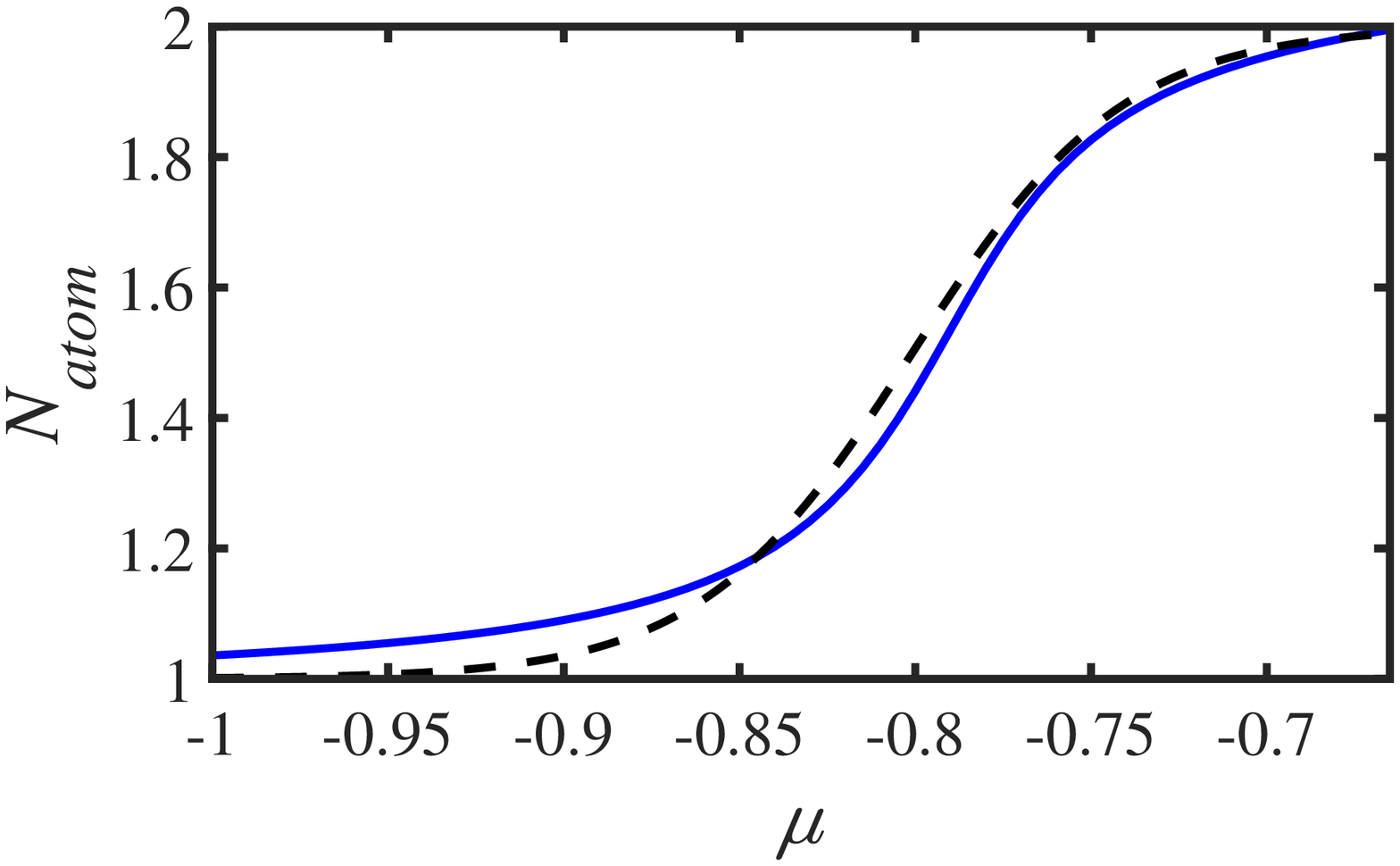}
  \end{tabular}
  \caption{The atomic fragment occupation numbers $N\uat$ (solid blue line) compared to the Fermi-Dirac $\bar{n}$ (dashed black line). Top: $T = 1050~K$ and $\varepsilon_{0}^{(0)} = -1.5586$. Bottom: $T = 9500~K$ and $\varepsilon_{1}^{(0)} = -0.8008$.}
\label{fig:FD}
\end{figure}

The smoothening of the $N$ vs. $\mu$ staircase in Fig.1 suggests a possible analogy between finite distances and finite temperatures.  
In Fig. \ref{fig:FD}, we compare our calculated $N\uat$ to the average number of particles $\bar{n}$ from a Fermi-Dirac (FD) distribution:
\begin{equation}
\bar{n} = \frac{1}{e^{(\varepsilon_{i}^{(0)} -\mu)/kT}+1}
\end{equation} 
where $k$ is Boltzmann's constant and $T$ is the temperature. It is apparent from the figures that the analogy is not exact. The FD distribution at specified (unphysical) temperatures can capture some of the behavior of $N\uat(\mu)$ around the step between integer numbers or the upper region of the curve as it flattens near the integer. It cannot capture both at once, or correctly follow the behavior of the lower region as it rises from the lower integer.
%

\section{Conclusions and Outlook}   %
\label{sec:outlook}

We have shown that the chemical potential of an integer-electron system can be smaller than $I-A$ when the system (here, an atom) is at interacting distances from a metallic reservoir of electrons.  A continuous change in a {\em global} molecular property, $\mu$, distorts the density of one fragment (either metal or atom) markedly more than the density of the other fragment.
The typical $N_{atom}$ vs. $\mu$ staircase function is smoothed-out as a result of the finite-distance interactions between the atom and the metal surface. Our method is useful for calculations on semi-infinite systems and allows treatment of different fragments with different computational techniques. For example, an atomic or a molecular fragment can be treated with an accurate wave-function method and the semi-infinite metal fragment can be treated with a more innate Green's function method. The method provides a convenient way to account for the finite-distance interactions near the metal surface. 

In the extension of Frozen-density embedding \cite{WW96} to fragments with non-integer particle numbers \cite{FLS14}, the total energy is minimized under the constraint that each fragment density integrates to a pre-established fractional value. In this method, each fragment has a different chemical potential along with a different embedding potential, and the fractional charges on the fragments are not an output but an input for the calculation. As an alternative, we have proposed {\em chemical-potential equalization} as the main criterion for determining fractional charges in density embedding. Because calculating fractional charges is important in various fields, from electrolysis \cite{L79,L85} to catalysis \cite{LKSTSTDJNMFMNL16}, solar cells and organic electronics \cite{HRSH15, LTS14}, we anticipate several potential uses of the proposed approach.

Next steps include the investigation of exchange-correlation effects, the application of chemical-potential constrained P-DFT to realistic systems of atoms and molecules adsorbed on metal surfaces, and the calculation of surface resonance lifetimes through complex scaling \cite{Moiseyev}. 


	

\bibliographystyle{tfo}
\bibliography{Kelsies-version}

\end{document}